# Coarse-grained graph architectures for all-atom force predictions

Sungwoo Kang[1,2,†,*] and Jinwoong Chae[1,†]

[1]Computational Science Research Center, Korea Institute of Science and Technology (KIST), Seoul 02792, Republic of Korea

[2]Division of Nanoscience and Technology, KIST School, University of Science and Technology (UST), Seoul 02792, Republic of Korea

†These authors contributed equally.

*Correspondence: sung.w.kang@kist.re.kr

We introduce a machine-learning framework termed coarse-grained all-atom force field (CGAA-FF), which incorporates coarse-grained message passing within an all-atom force field using equivariant nature of graph models. The CGAA-FF model employs grain embedding to encode atomistic coordinates into nodes representing grains rather than individual atoms, enabling predictions of both grain-level energies and atom-level forces. Tested on EC/EMC organic electrolytes and RDX crystalline and disordered phases, CGAA-FF achieves 0.201 and 0.253 eV Å$^{-1}$, respectively, while providing about 10-fold and 5-fold higher computational speed and memory efficiency, respectively, than conventional MLIPs. Since this CGAA framework can be integrated into any equivariant architecture, we believe this work opens the door to efficient all-atom simulations of soft-matter systems.

## Introduction

Recently, there have been active developments in machine-learning interatomic potentials (MLIPs), which are data-driven models trained using density-functional theory (DFT) calculations to predict energies and forces for atomistic configurations.[1–3] Despite their wide application in inorganic systems,[4–8] their use in soft materials such as organic liquids or polymers has remained limited. This limitation primarily arises from the large length and time scales required for simulations, as well as the numerous conformations of molecular units, significantly increasing the computational cost of generating training sets. Previous studies have focused on dataset sampling to improve the accuracy of intermolecular interactions;[9–15] however, few have addressed modifications to model architectures to increase computational cost.

Most MLIPs fall into two categories: descriptor-based and graph-based models.[16] Descriptor-based models exhibit rapidly increasing computational costs and decreasing accuracy and data efficiency with the number of elements,[3] making them unsuitable for organic systems typically containing at least five elements (e.g., C, H, O, N, S). Graph-based models exhibit computational costs independent of element number[17,18] and demonstrate better transferability.[19] For example, ref. 20 showed that the pretrained graph MLIP, SevenNet-0,[21] accurately simulates organic electrolytes with minimal fine-tuning. Nevertheless, graph-based models typically involve numerous parameters, leading to high computational costs and memory requirements, thus constraining simulation scales required for organic systems.

For a different direction, coarse-grained (CG) models simplify the representation of molecular systems by grouping several atoms into single interaction sites. Recent developments in CG models based on machine learning methods have made them much more flexible compared to conventional CG potentials based on simple physical models.[22–24] While CG models are suitable for modeling much larger systems, they sacrifice accuracy due to the loss of all-atom degrees of freedom. (However, an exception may arise when predicting coarse-scale properties, where CG models can outperform all-atom (AA) models; e.g., ref. 25.). Here, we develop a class of machine-learning CG energy models employing equivariant graph architectures, which also incorporates all-atom information to enable the prediction of all-atom forces.

## Results

The model predicts the total energy ($E_{\mathrm{tot}}$) as a sum of grain energies ($E_{\mathrm{grain}}$) as follows:

$$E_{\mathrm{tot}} = \sum_{J} E_{\mathrm{grain},J}\,, \tag{1}$$

where $J$ is the grain index. A grain is defined by $\mathbf{R}_J$ and $\mathbf{r'}_j$, where $\mathbf{R}_J$ is the center of grain, and $\mathbf{r'}_j$ is the relative position of an atom $j$ included in the grain. These can be written below:

$$\mathbf{R}_J = \frac{1}{N_J}\sum_{j} \mathbf{r}_j\,, \tag{2}$$

$$\mathbf{r'}_j = \mathbf{r}_j - \mathbf{R}_J, \tag{3}$$

where $N_J$ is the number of atoms included in the grain $J$, and $\mathbf{r}_j$ is the absolute coordinate of atom $j$. The use of the center of mass is also possible, but future studies are needed to assess whether the ML model can effectively learn atomic mass information. As illustrated in Fig. 1a, we construct the graph model architecture where the nodes are predefined types of grains. The model encodes $\mathbf{r'}_j$ as an array of vectors (Fig. 1b); the coordinates of each atom are assigned in a predefined order to the array components corresponding to their grain type, and the components associated with other grain types are set to $\mathbf{0}$. These arrays are concatenated with a 0e irreps (1 or 0 depending on grain

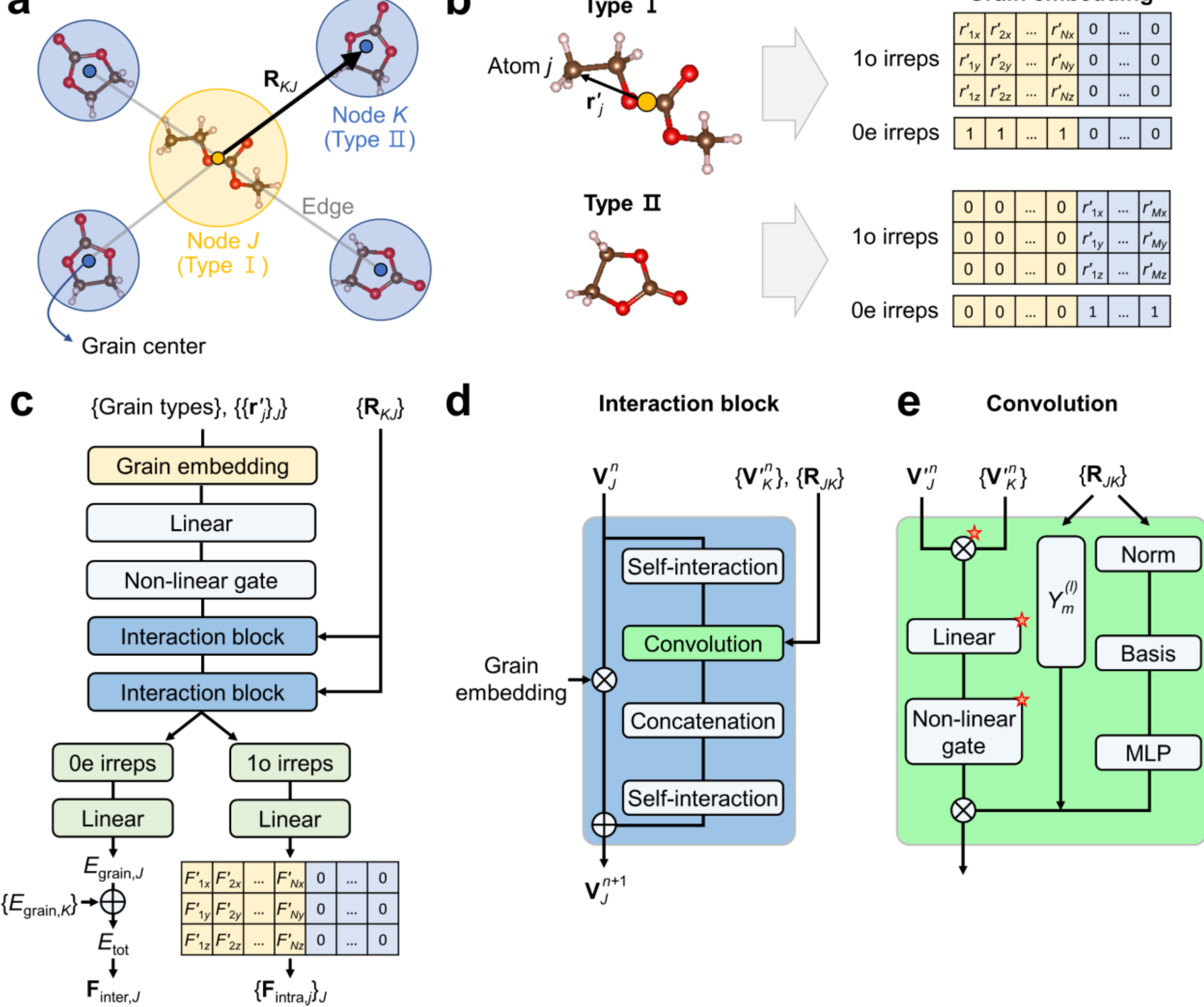

**Fig. 1 Schematic of the CGAA-FF architecture.** Illustration of the **a,** grain-based graph representation, **b,** node embedding. **c,** overall CGAA-FF model structure, **d,** interaction block, and **e,** convolution layers. The star marks in e indicate components that differ from the original NequIP model. Uppercase letters (e.g., $J$) denote grains, whereas lowercase letters (e.g., $j$) denote atoms. $N$ and $M$ indicate the numbers of atoms in the corresponding grains, and $R_{KJ}$ denotes the distance between the centers of grains $K$ and $J$. In $V_J^n$, $V$ represents the feature tensor, $J$ denotes the grain index, and $n$ indicates the interaction-layer index.

type) and used as an input feature. We refer to this process as grain embedding.

Fig. 1c illustrates the network architecture, where the NequIP is adopted as a base model.[17] After embedding, the number of features is altered through a linear operation (self-interaction) and then the nodes pass through non-linear gates. Unlike the original model, after processing through the final interaction blocks, the node value remains a combination of 0e and 1o irreducible representations (irreps). 0e irreps (scalars) are used to predict grain energies and 1o irreps (vectors) are used to calculate internal forces. Specifically, the atomic forces can be written as follows:

$$F_{j\alpha} = -\frac{\partial E_{\text{tot}}}{\partial r_{j\alpha}} = -\sum_{k\in\{n_J\}} \frac{\partial E_{\text{tot}}}{\partial r'_{k\alpha}}\frac{\partial r'_{k\alpha}}{\partial r_{j\alpha}} - \frac{\partial E_{\text{tot}}}{\partial R_{J\alpha}}\frac{\partial R_{J\alpha}}{\partial r_{j\alpha}}. \quad (4)$$

where $\alpha$ denotes the directional index ($x$, $y$, or $z$), and $n_J$ is the index for an atom in grain $J$. Using eqs. 2,3:

$$F_{j\alpha} = -\frac{\partial E_{\text{tot}}}{\partial r'_{j\alpha}} + \frac{1}{N_J}\sum_{k\in\{n_J\}} \frac{\partial E_{\text{tot}}}{\partial r'_{k\alpha}} - \frac{1}{N_J}\frac{\partial E_{\text{tot}}}{\partial R_{J\alpha}}. \quad (5)$$

We then define intra-grain forces ($\mathbf{F}_{\text{intra},j}$) and inter-grain forces ($\mathbf{F}_{\text{inter},J}$) as below:

$$\mathbf{F}_j = \mathbf{F}_{\text{intra},j} + \frac{1}{N_J}\mathbf{F}_{\text{inter},J}, \quad (6)$$

$$F_{\text{inter},J\alpha} = -\frac{\partial E_{\text{tot}}}{\partial R_{J\alpha}}. \quad (7)$$

$\mathbf{F}_{\text{inter},J}$ is calculated by differentiating $E_{\text{tot}}$ with respect to the inter-grain distances, as done in conventional MLIPs. $\mathbf{F}_{\text{inter},j}$ is directly predicted from the 1o irreps. Specifically, the intra-grain forces are computed using a final output block whose size is equal to the sum of $N_{\mathrm{atom}}$ over all grain types. For a given input configuration, only the intra-grain force components corresponding to the relevant grain type are output, while the remaining components are set to $\mathbf{0}$. From $\mathbf{F}_{\mathrm{intra},j}$, and $\mathbf{F}_{\mathrm{inter},J}$, atomic force $\mathbf{F}_j$ can be computed, enabling AA molecular dynamics (MD) simulations.

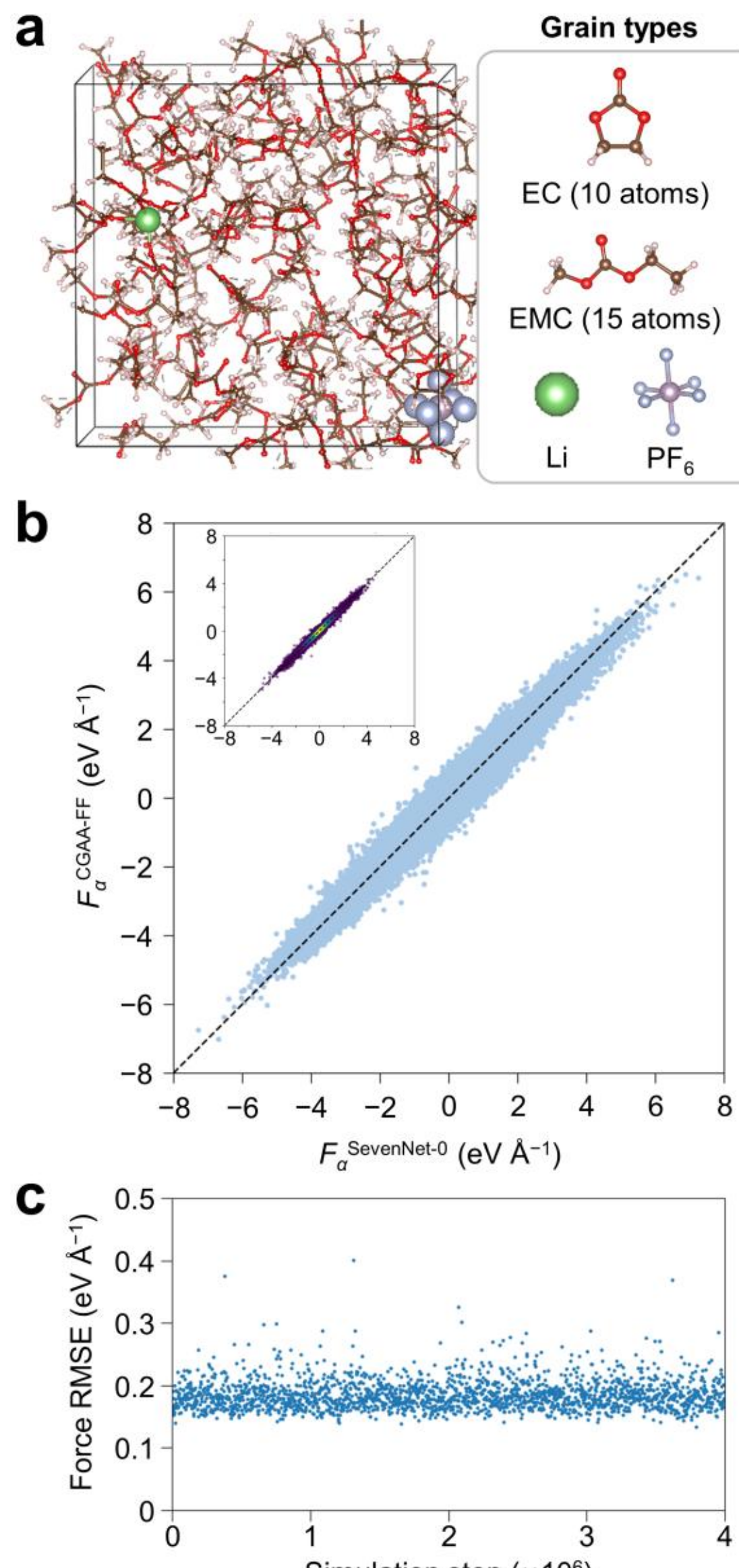

**Fig. 2. Test of the CGAA-FF model for the EC/EMC system. a,** Illustration of the test system. **b**, Parity plot for the force prediction between SevenNet-0 and CGAA-FF. $\alpha$ denotes the directional index ($x$, $y$, or $z$). Inset is the density-mapped parity plot for the randomly sampled data. **c**, Force RMSE of CGAA-FF relative to SevenNet-0 for configurations sampled during MD simulations performed using CGAA-FF.

Fig. 1d,e shows the structure of interaction blocks and convolution layers, respectively. While most of the architecture follows the NequIP model,[17] we modify the convolution layers to better capture inter-grain interactions. Each convolution layer involves tensor products between node features ($\mathbf{V'}_J^n$, where $n$: interaction block index) of neighboring grains, followed by self-interactions and non-linear gate activations (star marks in Fig. 1e, see Methods section for details).

As the model utilizes and predicts all-atom properties, it is technically not a CG model. Instead, it can be viewed as an AA force field in which message passing is performed at the CG level. Consequently, the number of edges is reduced by a factor of $N_J{}^2$, decreasing computational costs to $O(<N_J{}^2>^{-1})$. Since this model integrates the advantages of CG approach in an AA force field, we term this class of models coarse-grained all-atom force fields (CGAA-FF).

As the first test system, we model electrolytes composed of ethylene carbonate (EC) and ethyl methyl carbonate (EMC), combined with Li and $PF_6$ salts, defining each individual molecule or salt as a separate grain (Fig. 2a). This system is particularly challenging to model using conventional CG approaches due to the significant molecular anisotropy of EMC and large atom numbers in grains (EC: 10, EMC: 15). The training set is generated through MD simulations using SevenNet-0[21] with D3 van der Waals correction,[26] as SevenNet-0-D3 is known to accurately describe EC and EMC liquids even without fine-tuning.[20] The root-mean-square errors (RMSEs) of validation set for energy and force are 4.96 meV atom$^{-1}$ and 0.201 eV Å$^{-1}$, respectively. Fig. 2b presents the parity plot for the predicted force components. The force values range approximately ±7.5 eV Å$^{-1}$, indicating that the force error relative to the maximum force is around 2.7%. Given that this value is ~3% for the conventional MLIP with the similar system,[10] we demonstrate that our model achieves reasonable accuracy. We further confirm that the force RMSE remains at a similar level to the training error throughout the $4\times10^6$ steps (2 ns) of NVT simulations conducted using CGAA-FF (Fig. 2c).

**Table 1. Summary of CGAA-FF test results for the EC/EMC system.**

| | Model | CGAA-FF | SevenNet-bespoke | SevenNet-0 |
|---|---|---|---|---|
| Architecture | Cutoff (Å) | 10 | 5 | 5 |
| | Number of interaction layers | 2 | 5 | 5 |
| | Irreps | 64×e0 +64×o1 | 64×e0 +64×o1 | 128×e0 +64×e1 +32×e2 |
| Performance | Energy RMSE (meV atom$^{-1}$) | 4.96 | 6.7 | - |
| | Force RMSE (eV Å$^{-1}$) | 0.201 | 0.044 | - |
| | Computational cost [h (1000 atom)$^{-1}$ ns$^{-1}$] | 4.4 | 33.3 | 96.0 |
| | Memory (MB atom$^{-1}$) | 0.49 | 2.02 | 6.82 |

Regarding computational speed, SevenNet-0 requires 0.17 s per simulation step for 983-atom cell on a single NVIDIA A6000 GPU, corresponding to 96.1 h (1000 atom)$^{-1}$ ns$^{-1}$. CGAA-FF takes 4.4 h (1000 atom)$^{-1}$ ns$^{-1}$, approximately 22 times faster than SevenNet-0. Regarding memory usage, SevenNet-0 consumes 5248 MB for a 983-atom simulation, whereas CGAA-FF requires 10,331 MB for a 26,541-atom simulation, representing roughly 14 times less memory per atom compared to SevenNet-0. Since the computational cost can also be reduced by decreasing the hyperparameters of conventional MLIP architectures, we train a bespoke SevenNet model with a similar model size and cutoff radius to CGAA-FF. As shown in Table 1, the bespoke SevenNet model achieves higher force accuracy than CGAA-FF, but at the cost of lower computational speed and memory efficiency.

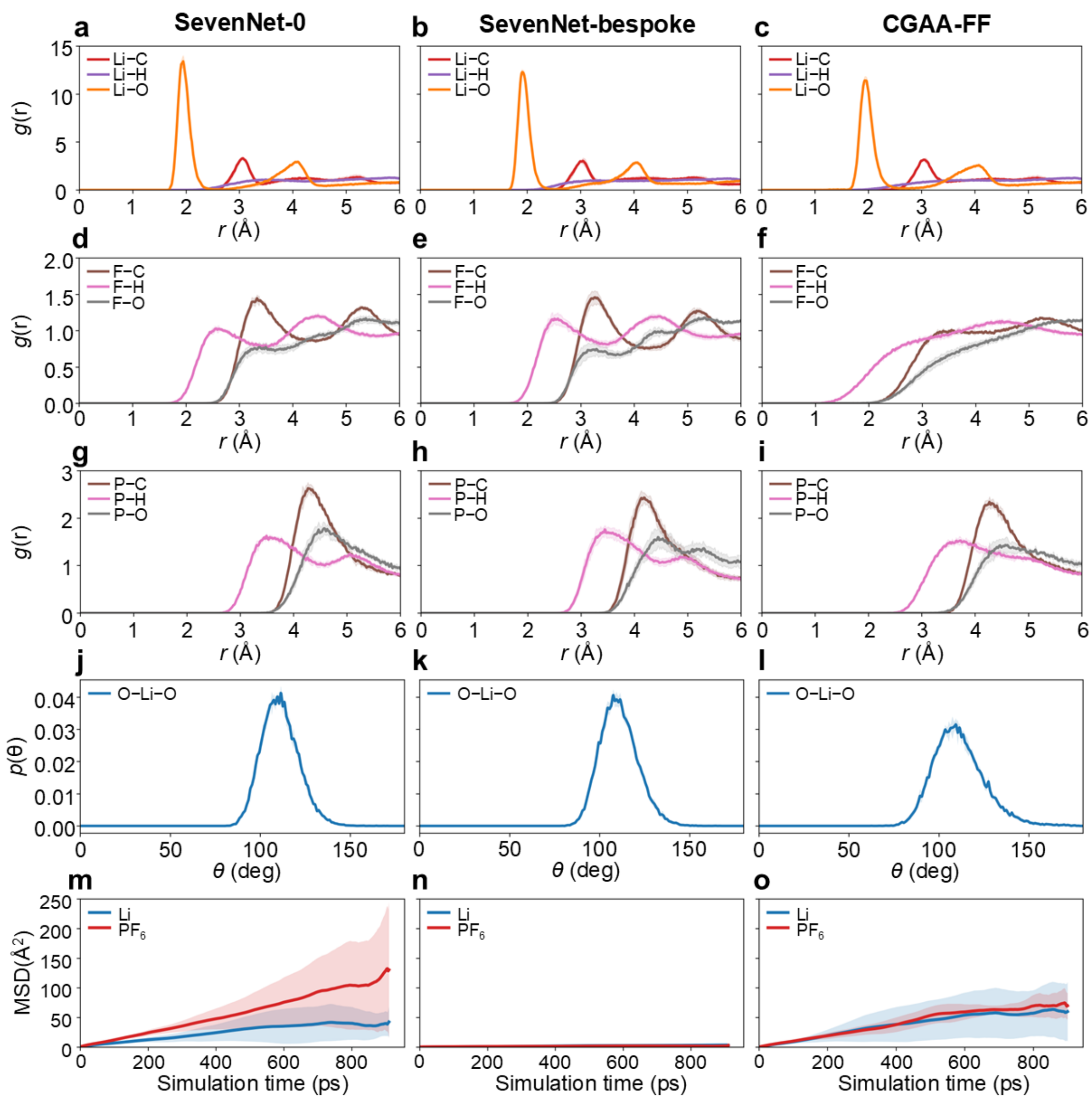

**Fig.3 Evaluation of structural and dynamics properties. a-c,** RDFs between Li and atoms in the electrolyte molecules, **d-f,** RDFs between F and atoms in the electrolyte molecules, **g-i,** RDFs between Li and atoms in the electrolyte molecules, **j-l,** ADFs between Li and O, and **m-o,** MSDs over simulation times of SevenNet-0, SevenNet-bespoke, and CGAA-FF models, respectively.

We also evaluate the radial distribution functions (RDFs), angular distribution functions (ADFs), and diffusion properties for SevenNet-0, SevenNet-bespoke, and CGAA-FF models (Fig. 3). For the RDFs between Li and atoms in the electrolyte molecules (Fig. R3a–c), the three models show similar trends. However, the RDFs between F ions and atoms in the electrolyte molecules (Fig. R3d–f) show broadened and less pronounced peaks in CGAA-FF. In addition, the ADFs of O-Li-O (Fig. 3j–l) indicates that CGAA-FF exhibits a broader angular distribution. The RDFs involving P and atoms in the electrolyte molecules (Fig. 3g–i) show relatively better agreement with the reference. These results suggest that, although CGAA-FF retains all-atom information, its network architecture tends to emphasize coarse-scale properties during learning. Consistently, the mean-square-displacement (MSD) results (Fig. 3m–o) show that CGAA-FF captures the diffusion behavior more accurately than the bespoke model, for which almost no diffusion is observed. Since the bespoke SevenNet model and CGAA-FF are trained on the same dataset, this result suggests that CGAA-FF may learn coarse-scale dynamical properties more effectively than all-atom MLIPs. We note that the equilibrium density of EC/EMC with Li and $PF_6$ salts evaluated by SevenNet-0, SevenNet-bespoke, and CGAA-FF are 1.24, 1.44 and 1.33 g cm$^{-3}$, respectively.

**Table 2. Summary of CGAA-FF test results for the RDX system.**

| Model | | CGAA-FF Type I | CGAA-FF Type II | SevenNet-bespoke | SevenNet-Omni |
|---|---|---|---|---|---|
| Architecture | Cutoff (Å) | 14 | 6 | 6 | 6 |
| | Number of interaction layers | 2 | 2 | 5 | 5 |
| | Irreps | 64×e0 +64×o1 | 32×e0 +32×o1 | 64×e0 +64×o1 | 128×e0 +128×o0 +64×e1 +64×o1 +32×e2 +32×o2 +32×e3 +32×o3 |
| Performance | Energy RMSE (meV atom$^{-1}$) | 1.95 | 5.45 | 2.76 | - |
| | Force RMSE (eV Å$^{-1}$) | 0.273 | 0.253 | 0.091 | - |
| | Cost [h (1000 atom)$^{-1}$ ns$^{-1}$] | 5.14 | 3.92 | 52.9 | 96.8 (Accelerated by FlashTP about 5 times) |
| | Memory (MB atom$^{-1}$) | 0.54 | 0.22 | 3.22 | 5.51 (Reduced by FlashTP about 5 times) |

For the second test system, we train a CGAA-FF model on crystalline and disordered structures of RDX molecules. This system was studied in ref. 24 using a graph-based CG model, allowing us to compare CGAA-FF with an existing CG approach. We construct the training set from MD simulations of crystalline RDX at 0–20 GPa and 250–800 K (Fig. 4a) using SevenNet-Omni.[27] We train two models with different grain assignments (Fig. R4b). In the first model (type I), the grain was defined at the whole-molecule level, such that one entire RDX molecule was treated as a single grain. In the second model (type II), the grain was defined using $NO_2$ and $C_3N_3H_6$ groups, where one RDX molecule consists of three $NO_2$ grains and one $C_3N_3H_6$ grain, which is identical to that used in ref. 24. Figs. 4c,d compare the predicted and reference forces evaluated on the validation dataset for the models trained with grain types I and II, respectively. The RMSEs for energy and force are 1.95 meV atom$^{-1}$ and 0.273 eV Å$^{-1}$ for the type I model, and 5.45 meV atom$^{-1}$ and 0.253 eV Å$^{-1}$ for the type II model, respectively (Table 2). The computational costs are 5.14 and 3.92 h (1000 atom)$^{-1}$ ns$^{-1}$ for the type I and type II models, respectively. The accuracies and computational costs of the two models are therefore comparable. However, the hyperparameters used for the type I model are substantially more demanding than those used for the type II model. These results indicate that, within the similar computational setting, the whole-molecule scheme is less accurate, but computationally efficient than smaller-grain scheme.

We then compare the computational cost of CGAA-FF with that of the ML-based CG model in ref. 24. Since that study did not report the speed of their code, we directly evaluated the computational cost using their publicly available code. The computational time of model of ref. 24 is 1.1 h (1000 atom)$^{-1}$ ns$^{-1}$ (1 fs time step), whereas the computational times of CGAA-FF are 5.14 and 3.92 h (1000 atom)$^{-1}$ ns$^{-1}$ for the type I and type II models (0.5 fs time step), respectively, on an NVIDIA A6000 GPU. Considering this difference in time step, the CGAA-FF models are only about two times slower for inference.

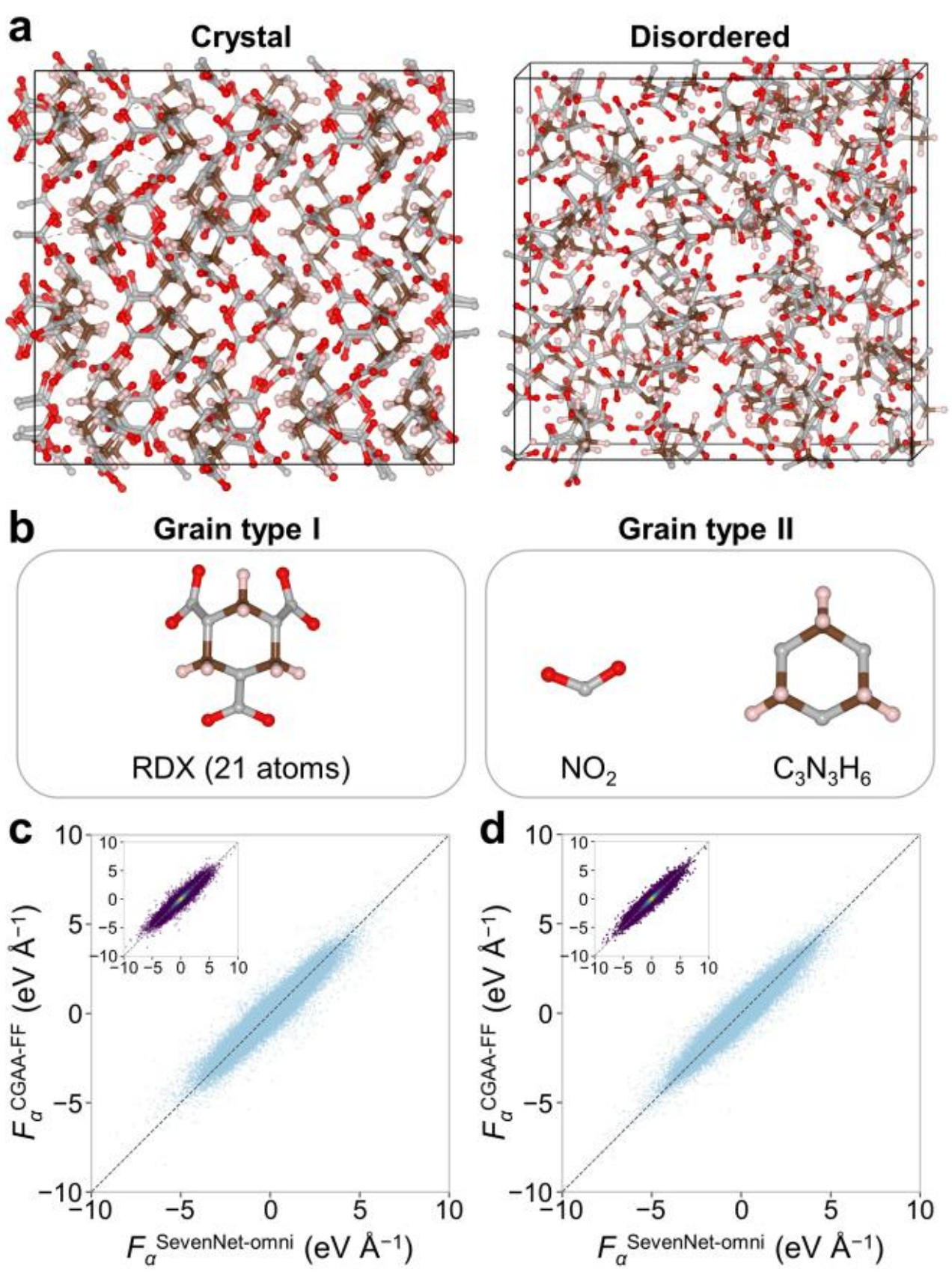

**Fig. 4. Test of CGAA-FF on the RDX system.** Schematic illustration of the **a,** test systems and **b,** grain types. Parity plot for the force prediction between SevenNet-0 and CGAA-FF for **c,** type I and **d,** type II models. $\alpha$ denotes the directional index ($x$, $y$, or $z$). Inset is the density-mapped parity plot for the randomly sampled data.

## Discussions

As this study represents the first implementation of the CGAA framework and given that the scheme is compatible with any equivariant model (just by replacing embedding and output blocks), we anticipate substantial performance improvements through future optimizations. For instance, we observe that introducing additional inter-grain interactions in convolutions (star marks in Fig. 1e) increase computational costs by approximately four times; thus, optimizing this part would greatly enhance inference speed. Additionally, choosing a more suitable base model and finely tuning hyperparameters could further improve computational efficiency.

Throughout the study, we use universal MLIPs for generating reference datasets. This is because, the conventional training of MLIPs for soft-matter systems require high computational cost for sampling configurations and force calculations for generating training set.[9–15] While pretrained MLIPs are known to suffer from softening of the potential energy surface (PES), this limitation can be efficiently solved by forgetting-aware fine-tuning strategies.[28] We further note that recently developed multi-domain universal MLIPs, such as SevenNet-Omni,[27] are jointly trained on inorganic and organic databases, and therefore show much better accuracy and less PES softening in soft-matter systems.

Note that the implementation of CGAA-FF does not satisfy energy conservation because intra-grain forces are not directly obtained from the gradient of the total energy. Consequently, both the energy and temperature show significant drift in NVE simulations (300 K → 700 K during 6000 steps), as also reported in the previous study.[29] In principle, an energy-conserving model within the CGAA-FF framework can be constructed by computing intra-grain forces using the following equation instead of direct prediction:

$$F_{\mathrm{intra},j\alpha} = -\frac{\partial E_{\mathrm{tot}}}{\partial r'_{j\alpha}} + \frac{1}{N_J}\sum_{k\in\{n_J\}}\frac{\partial E_{\mathrm{tot}}}{\partial r'_{k\alpha}}, \tag{8}$$

However, our tests show that this approach results in higher errors (>0.4 eV Å$^{-1}$) and unstable MD simulations. More advanced models and embedding methods will be explored in future work to achieve energy conservation. Additionally, we test a model that directly predicts total forces from 1o irreps without differentiation; this approach also leads to instability during MD simulations.

We note that a couple of previous studies have proposed similar concepts. The separation of intra- and inter-grain forces has been implemented previously in a covariant CG model.[24] However, this approach does not incorporate individual atomic information. Additionally, an energy-mapping strategy at the grain (termed 'monomer' in the literature) level was introduced in ref. [30], employing invariant intra- and inter-grain descriptors as inputs of neural network. Nevertheless, since the number of descriptors scales with O(<$N_J^2$>), computational efficiency and accuracy would decrease for systems with larger molecules. Consequently, it remains unclear whether this approach is feasible for large-molecule systems, as also acknowledged in the original study. In contrast, our work provides a comprehensive equivariant framework for grain-based representations, demonstrating its effectiveness for molecules containing up to 15 atoms. Therefore, although our method would incur higher computational costs due to equivariant constraints, it is complementary to aforementioned approaches, particularly advantageous for efficiently simulating larger molecules at the all-atom scale. For the same reason, however, we also note that the CGAA-FF model can be less effective for small-molecule systems.

Finally, we note that the CGAA-FF model has a limitation in that it cannot be directly used to train universal models. This is because the embedding layers scale with the number of grain types in the system, making the model prohibitively expensive for highly diverse chemical spaces. However, the main idea of our framework is to use vector quantities as features that represent the molecular geometry. In this sense, a related algorithm applicable to more general molecules is available in principle. One possible direction toward a more universal representation would be to avoid explicit enumeration of atomic vectors and instead construct features by pooling atomic vectors according to atom type (e.g., sum pooling). In such a representation, all atoms in a molecule could be encoded into a fixed-size embedding vector, making the model more transferable across different molecular species. The efficiency and accuracy of such an approach would require a thorough future study.

In summary, we have developed a theoretical framework that incorporates coarse-grained message passing for all-atom force fields, exploiting the equivariant nature of the recently developed graph-network architectures. The model demonstrates reasonable training and simulation accuracies and achieves an order-of-magnitude improvement in simulation speed and memory efficiency. Since the proposed scheme can be integrated into any equivariant model, further performance enhancements are expected in future studies. Thus, we believe this work paves the way towards fully atomistic modeling of soft-matter systems.

## Methods

**Training set generation.** MD simulations for generating the training dataset of EC/EMC system are performed using SevenNet-0[21] combined with D3 van der Waals (vdW) corrections[26] using the LAMMPS code.[31] The main configuration consists of 24 EC, 49 EMC, 1 Li, and 1 $PF_6$ molecules/salts. Separate NPT simulations are conducted independently at temperatures of 300, 350, 400, and 500 K for 500 ps each. NVT simulations are also independently performed at these same temperatures for 250 ps, using a fixed cell size of 21.5 Å × 21.5 Å × 21.5 Å, corresponding to the average volume from the 300 K NPT simulation. Additionally, MD simulations are carried out for three smaller configurations: one with two EC molecules, another with two EMC molecules, and a third containing one EC and one EMC molecule. These simulations use a cell size of 22.1 Å × 22.1 Å × 22.1 Å and are performed at 300 and 400 K under NVT conditions for 450 ps.

For RDX system, SevenNet-Omni[27] (*mpa* task) combined with D3 vdW corrections[26] are employed. 2 × 2× 2 supercells are employed. For crystalline phase, the 10 ps of NPT and 100ps of NVT simulations are performed at six conditions: (0 GPa, 250 K), (0 GPa, 300 K), (0 GPa, 500 K), (0 GPa, 800 K), (10 GPa, 300 K), and (20 GPa, 300 K). We also perform simulations by linearly varying temperature from 300 K to 800 K at 0 GPa.

For both systems, the timestep is set to 0.5 fs and the training configurations are sampled every 50 fs. For both NVT and NPT simulations, Nosé–Hoover algorithm is used with the damping parameter as 1 ps.

**Details of the CGAA-FF model.** The convolution operation $\mathcal{L}$ is expressed as follows:

$$\mathbf{V}''^{n}_{JK} = \boldsymbol{\sigma}\left(\mathbf{self}\left(\mathbf{V}'^{n}_{J} \otimes \mathbf{V}'^{n}_{K}\right)\right), \tag{9}$$

$$\mathcal{L}_{J,c,m_o}^{l_o,p_o,l_f,p_f l_i,p_i} = \sum_{m_f,m_i} C_{l_f,m_f l_i,m_i}^{l_o,m_o} \sum_{k\in S} \left( R\left(R_{JK}\right)_{c,l_o,p_o,l_f,p_f l_i,p_i} \right) Y_{m_f}^{l_f}\left(\hat{\mathbf{R}}_{JK}\right) V''^{l_i,p_i}_{JK,c,m_i}, \tag{10}$$

where, **self**$(\cdot)$ and $\boldsymbol{\sigma}(\cdot)$ represent the self-interaction function and non-linear gate activation, respectively. The $c$, $l$, $p$, and $m$ denote the feature index, rotation, parity, and projection order, respectively. $i$, $o$, and $f$ refer to input, output, and filter indices. $C$, $R$, and $Y$ denotes the radial function, Clebsch–Gordan coefficients, and spherical harmonics.

For the calculation of intra-grain forces, since the sum of $\mathbf{F}_{\text{intra},j}$ within a grain is $\mathbf{0}$, we subtract average force value from the predicted forces and the components associated with other grain types are set to $\mathbf{0}$ (see Fig. 2c).

**Model training.** We modify the SevenNet code[21] to construct the CGAA-FF model, so the input settings and hyperparameters follow the SevenNet format. For the radial function, we use neural networks composed of eight Bessel basis functions at the input layer followed by two hidden layers. The initial learning rate is set to 0.005 with an exponential scheduler, though we manually adjust it during training according to the observed learning curve. The model is trained simultaneously on energy, inter-grain force, and intra-grain force, with respective loss weights of 1.0, 1.0, and 20.0. We randomly split the data into training and validation sets at a 9:1 ratio.

**Test simulations.** We develop the ASE simulator[32] for CGAA-FF model. The simulations are conducted in 21.5 Å × 21.5 Å × 21.5 Å simulation cell which includes 24 EC, 74 EMC, 1 Li, and 1 $PF_6$ molecules/salts. The NVT simulations are performed with Langevin algorithm with the damping parameter of 50 fs. Configurations are sampled every 1 ps to estimate the error between SevenNet-0 and CGAA-FF, as shown in Fig. 2c.

**Simulation speed and memory usage measurement.** For simulation speed, we measure only the inference time and exclude the preprocessing step. This is because, in simulations using the ASE code, most of the computation time is spent on preprocessing—particularly in calculating grain centers under periodic boundary conditions and storing them in predefined 1o irreps arrays. Although this preprocessing is currently required at every simulation step when using the ASE calculator, it only needs to be done once if an optimized simulator for the CGAA-FF model is developed. This is because updates to atomic coordinates can be handled by simply modifying $\mathbf{R}_J$ and $\mathbf{r}'j$. Such an optimized implementation is planned for future work. For memory usage measurement, we use the max_memory_allocated() function in torch.cuda module. The speed and memory of SevenNet-0 and CGAA-FF are measured using the simulation cells consisting of 983 and 26,541 atoms (3×3×3 supercell), respectively. For RDX system, 2×2×2 supercells (1,344 atoms) are used for evaluations.

**Visualization of atomic structures.** All atomic configurations in this paper are drawn with VESTA code.[33]

## Data availability

The training data, input, and trained models are available at https://doi.org/10.6084/m9.figshare.28914470.v1.

## Code availability

The code used in the current study is available at https://github.com/kang1717/CGAANet_proto.

## Acknowledgement

This work was supported by the Nano and Material Technology Development Programs through the National Research Foundation of Korea (NRF) funded by the Ministry of Science and ICT (Grant No. RS-2024-00407995 and No. RS-2024-00450102).

## Competing interests

The authors declare no competing interests.